\documentclass[twocolumn,showpacs,prd]{revtex4}

\newcommand{\beq}{\begin{equation}}
\newcommand{\eeq}{\end{equation}}
\newcommand{\beqar}{\begin{eqnarray}}
\newcommand{\eeqar}{\end{eqnarray}}

\newcommand{\del}{\partial}
\newcommand{\dell}[1]{\stackrel{\leftarrow}{\partial^{#1}}}
\newcommand{\dells}[1]{\stackrel{\leftarrow}{\partial_{#1}}}
\newcommand{\delr}[1]{\stackrel{\rightarrow}{\partial^{#1}}}
\newcommand{\dellr}[1]{\stackrel{\leftrightarrow}{\partial^{#1}}}
\newcommand{\dellrs}[1]{\stackrel{\leftrightarrow}{\partial_{#1}}}
\newcommand{\dellrss}[2]{\stackrel{\leftrightarrow}{\partial_{#1}^{#2}}}
\newcommand{\dellss}[2]{\stackrel{\leftarrow}{\partial_{#1}^{#2}}}
\newcommand{\delrss}[2]{\stackrel{\rightarrow}{\partial_{#1}^{#2}}}

\ifx\undefined\fiverm
     \font\fiverm=cmr5
\fi

\input{prepictex}
\input{pictex}
\input{postpictex}

\begin{document}
\title{Quantum Field Theory without Infinite Renormalization}
\author{Tarun Biswas}
\email{biswast@newpaltz.edu}
\affiliation{State University of New York at New Paltz, \\ New Paltz,  NY 12561, USA.}
\date{\today}

\begin{abstract}
Although Quantum field theory has been very successful in explaining experiment, there are
two aspects of the theory that remain quite troubling. One is the no-interaction result
proved in Haag's theorem. The other is the existence of infinite perturbation expansion
terms that need to be absorbed into theoretically unknown but experimentally measurable quantities
like charge and mass -- i.e. renormalization. Here it will be shown that the two problems
may be related. A ``natural'' method of eliminating the renormalization problem also sidesteps
Haag's theorem automatically. Existing renormalization schemes can at best be considered a
temporary fix as perturbation theory assumes expansion terms to be ``small'' -- and
infinite terms are definitely not so (even if they are renormalized away). String theories  may
be expected to help the situation because the infinities can be traced to the point-nature of
particles. However, string theories have their own problems arising from the extra space dimensions required. Here a more directly physical remedy is suggested. Particles
are modeled as extended objects (like strings). But, unlike strings, they are 
composites of a finite number of constituents each of which resides in the normal
4-dimensional space-time. The constituents are bound together by a manifestly
covariant confining potential. This approach no longer requires infinite
renormalizations. At the same time it sidesteps the no-interaction result proved 
in Haag's theorem.
\end{abstract}
\pacs{03.70.+k, 11.10.-z, 11.10.Cd, 12.20.-m}
\maketitle

\section{Introduction}
The infinities in perturbative Quantum Field Theories (QFT) can be traced to the
point-nature of particles. If particles could be modeled as extended
objects such infinities might disappear. String theories consider particles to be extended
objects and hence they might solve the problem. However, string theories have problems
of their own (for example, extra space dimensions). Renormalization schemes have worked,
but the mathematics of a perturbation theory with expansion terms that are {\em infinite}
is not satisfactory -- even if these terms are absorbed into experimentally measured 
quantities like charge and mass. Here another attempt
is made to model particles as extended objects -- they are seen as composites of a finite
number of constituent point objects. The fact that the constituents are once
again point objects does not recreate the original problem of infinite
perturbative terms because of some additional caveats that will be discussed
soon. Each constituent resides in the usual physical 4-dimensional space-time.
Composites of interacting point objects were considered to be inconsistent
with relativity at one time\cite{sudarshan}. However, later, more sophisticated
mathematical techniques were discovered to deal with such composites in a relativistically
covariant form\cite{arens,droz,todorov,komar,biswas1}. A quantum field theory
of such composites has been used to model hadrons as composites of quarks\cite{biswas2}.
Here a similar formulation will be used to create composite models for particles
that are traditionally not considered to be composites (electrons, photons, etc.).

The proposed composite model of a particle has one {\em vertex} component with one or 
two {\em satellite} components. The
vertex and the satellites are interacting point objects. A quantum field theory of
such composites is a second quantized theory but has only first quantized interactions 
between satellites and the vertex. This mix of first and second quantization can be
achieved quite seamlessly\cite{biswas2}. Some mathematical similarities exist between
this model and string theories. However, this theory is significantly easier to visualize
as it does not require objects to reside in anything other than normal 4-dimensional
space-time. The vertex and satellites are somewhat like
virtual particles of standard QFT. However, not being individually
second quantized, they do not produce the ill-effects of an infinite number of them
being created internally.

Experimentally observed effects like anomalous magnetic moment of electrons and the
Lamb shift can be explained effectively by the dynamics of the satellites
within particles. Hence, there is no need for closed loop diagrams of the particles
themselves. As the closed loop diagrams are responsible for the offending infinities,
eliminating them would eliminate the infinities. However, such diagrams are a
natural consequence of perturbation theory. They appear due to the assumption that
the free fields and the interacting fields are related by some unitary operator
$U_{I}$. Now, the same assumption also leads to the no-interaction theorem
of Haag\cite{wightman}. So it is natural to suspect the unitarity of $U_{I}$. If
$U_{I}$ were not unitary, it would solve two problems at once. Here it is proposed
that the effective part of $U_{I}$ (call it $V_{I}$) be obtained by removing 
from $U_{I}$ all contributions from closed loop diagrams. This will get rid of all
inifinities and at the same time render $V_{I}$ nonunitary. Hence, the no-interaction
problem of Haag will also be solved.

\section{The structure of bosons}
\label{boson}
\begin{figure}
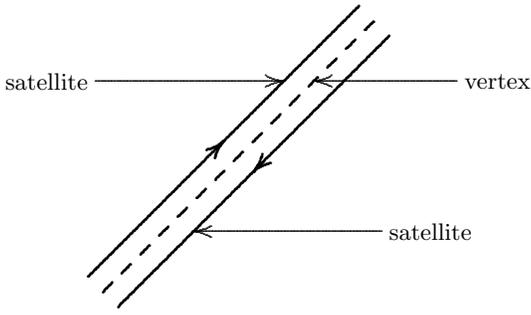

\beginpicture
\setcoordinatesystem units <1mm,1mm>
\setplotarea x from -25 to 25, y from -20 to 20
\setplotsymbol ({.})
\setdashes
\plot -18 -18  18 18 /
\setsolid
\arrow <7pt> [.2,.67] from -20 -16 to -2 2
\plot -2 2  16 20 /
\arrow <7pt> [.2,.67] from 20 16 to 2 -2
\plot 2 -2  -16 -20 /
\setplotsymbol ({\fiverm .})
\put {satellite} [r] at -20 10
\arrow <7pt> [.2,.67] from -19 10 to 6 10
\put {satellite} [l] at 20 -10
\arrow <7pt> [.2,.67] from 19 -10 to -6 -10
\put {vertex} [l] at 30 10
\arrow <7pt> [.2,.67] from 29 10 to 10 10
\endpicture
\caption{Model of a boson \label{fig1}}
\end{figure}
In the present model, a bosonic particle is constituted of one spin-zero vertex
and two spin-half satellites. In Feynman diagrams it will be represented as
shown in figure~\ref{fig1}. The dashed line represents the vertex and the satellites
are represented by solid lines with arrows. The arrows suggest the fermion and
anti-fermion nature of the satellites. The position and momentum four-vectors of
the vertex are $q_{0}$ and $p_{0}$\footnote{the four-vector indices are suppressed for
compactness.}. The positions and momenta of the satellites are
$q_{a}$ and $p_{a}$ ($a=1,2$). It is convenient to use coordinates relative to the
vertex position and hence we define the following.
\beqar
P & = & p_{0}+p_{1}+p_{2}, \nonumber \\
\pi_{1} & = & p_{1}, \nonumber \\
\pi_{2} & = & p_{2},\nonumber \\
Q & = & q_{0}, \nonumber \\
\xi_{1} & = & q_{1}-q_{0}, \nonumber \\
\xi_{2} & = & q_{2}-q_{0}. \label{eq1}
\eeqar
The resulting 24 dimensional phase space can be
represented as
\beq
S_{p1}= \{P,\pi_{1},\pi_{2},Q,\xi_{1},\xi_{2}\}, \label{eq2}
\eeq
where the only nonzero commutators are
\beq
[Q,P]=i\eta,\;[\xi_{1},\pi_{1}]=i\eta,\;[\xi_{2},\pi_{2}]=i\eta; \label{eq3}
\eeq
where $\eta$ is the metric with signature $(-+++)$ and units are chosen
such that $\hbar=1$.
It has been shown earlier\cite{biswas2} that $S_{p1}$ can be canonically transformed to $S_{p}$
given by
\beq
S_{p}= \{P,\pi_{1}^{\|},\pi_{1}^{\bot},\pi_{2}^{\|},\pi_{2}^{\bot},x,
\xi_{1}^{\|},\xi_{1}^{\bot},\xi_{2}^{\|},\xi_{2}^{\bot}\}, \label{eq4}
\eeq
where $P$ and $x$ are four-vectors representing the momentum and position of
the particle as a whole. $\pi_{a}^{\|}$ and $\xi_{a}^{\|}$
are components of $\pi_{a}$ and $\xi_{a}$ parallel to $P$ and
$\pi_{a}^{\bot}$ and $\xi_{a}^{\bot}$ are projections of $\pi_{a}$ and $\xi_{a}$
orthogonal to $P$. These components are obtained through the following
projection operators.
\beq
-\hat{P}= -P/\sqrt{-P^{2}},\;\; P^{\bot}=\eta + \hat{P}\hat{P}, \label{eq5}
\eeq
where $P^{2}=P\!\cdot\! P$ and ``$\cdot$'' represents the usual four vector
``dot'' product. Note that $\hat{P}$ has one suppressed index and $P^{\bot}$
has two suppressed indices -- the product in the definition of $P^{\bot}$ is a tensor product.
Then for any four-vector $v$ we define
\beq
v^{\|}=-v\!\cdot\!\hat{P},\;\;v^{\bot}=v\!\cdot\! P^{\bot}. \label{eq6}
\eeq
$v^{\|}$ is a scalar that is the zeroth component of $v$ in the center of mass (CM)
frame and $v^{\bot}$ is a four-vector with no zeroth component in the CM frame. Hence,
$v^{\bot}$ is effectively a three-vector in the CM frame.
It is to be noted that $x$ is {\em not} the position of the vertex\cite{biswas2}.
It is defined to remedy the problem of $Q$ having complicated commutators with
some of the other members of $S_{p}$. The result is the following set of
non-zero commutators in $S_{p}$.
\beq
[x,P]=i\eta,\;\;[\xi_{a}^{\|},\pi_{a}^{\|}]=-i,\;\;
[\xi_{a}^{\bot},\pi_{a}^{\bot}]=iP^{\bot}. \label{eq7}
\eeq
A maximal mutually commuting subset of $S_{p}$ is
\beq
S_{L}= \{x,\xi_{1}^{\|},\xi_{1}^{\bot},\xi_{2}^{\|},\xi_{2}^{\bot}\}. \label{eq8}
\eeq
Hence, the first quantized wavefunction of the system can be
written as a function on $S_{L}$.
\beq
\psi = \psi(x,\xi_{1}^{\|},\xi_{1}^{\bot},\xi_{2}^{\|},\xi_{2}^{\bot}). \label{eq9}
\eeq
The commutation conditions of equation~\ref{eq7} lead to the following
differential operator representation of the momenta.
\beq
i(-\pi_{a}^{\|},\pi_{a}^{\bot})=\del_{a\alpha}\equiv
\left(\frac{\del}{\del\xi_{a}^{\|}},\nabla_{a}\right), \label{eq10}
\eeq
and
\beq
iP_{\alpha}=\del_{\alpha}\equiv\frac{\del}{\del x^{\alpha}}, \label{eq11}
\eeq
where $\alpha$ is the four-vector index and $a$ the satellite number. The
four-vector component notation $(\cdots,\cdots)$ gives the zeroth component
as the first argument and the three-vector components as the second argument.
The three-vector operator $\nabla_{a}$ is defined to be the gradient in the 
three-vector space of $\xi_{a}^{\bot}$.
\beq
\nabla_{a}\equiv \left(\frac{\del}{\del\xi_{a1}},\frac{\del}{\del\xi_{a2}},
\frac{\del}{\del\xi_{a3}}\right). \label{eq12}
\eeq
The wavefunction $\psi$ must satisfy an equation of motion for each of the
satellites and one for the vertex. The one for the vertex can be replaced by
the whole particle equation. Hence, for free fields, it will be as follows.
\beq
D_{0}\psi\equiv \left(\del_{\mu}\del^{\mu}-m^{2}-\mu^{2}\right)\psi =0. \label{eq13}
\eeq
Here, $m$ is the total rest mass of the particle. It includes all internal energies.
The interaction energies of the satellites and the vertex are not physically
separable. Hence, for mathematical convenience, an arbitrary separation is made
such that the vertex energy in the CM frame is chosen as
zero. Thus, the CM energy of the particle will be the sum of the CM energies
of the two satellites alone. These can be seen to be the $\pi_{a}^{\|}$'s. However,
the rest mass $m$ of the particle {\em is} its CM energy. So,
\beq
m=\pi_{1}^{\|}+\pi_{2}^{\|}. \label{eq14}
\eeq
$\mu^{2}$ is a constant parameter that physically tends to zero. However, it is needed in
situations where $m=0$. In such situations $P^{2}$ would be zero and hence $\hat{P}$
would be ill-defined unless a positive $\mu^{2}$ is included.

The equations of motion of the satellites cannot be expected to be free field
equations. However, they can be made to {\em appear} like free spin-half fermion equations
by including an interaction function in the mass term\footnote{This is the conceptual
equivalent of the one-dimensional wave equation of string theories. However, it is
more general, as it allows for a large class of binding potentials in three space
dimensions. String theories
effectively use only the one-dimensinal particle in a box potential.}. So, we write
\beq
D_{a}\psi\equiv (\gamma_{a}^{\alpha}\del_{a\alpha}+m_{a})\psi = 0, \label{eq15}
\eeq
where $a=$1 or 2 for the two satellites. $\gamma_{a}^{\alpha}$ is a Dirac
matrix that operates on the sector of the wavefunction that corresponds
to satellite number $a$. $m_{a}$ is, in general, a function of all translation
invariant members of $S_{p}$ which excludes only $x$. However, further
conditions must be imposed on $m_{a}$ to maintain the consistency of the three
equations of motion. This means the three operators $D_{0}$, $D_{1}$ and
$D_{2}$ must obey the following conditions.
\beq
[D_{A}D_{B}-D_{B}D_{A}]\psi=0,\;\;\;\; A,B=0,1,2. \label{eq16}
\eeq
The simplest non-trivial way of satisfying equation~\ref{eq16} is to have
$m_{a}$ to be a function of $\xi_{a}^{\bot}$ alone.
\beq
m_{a}=m_{a}(\xi_{a}^{\bot}). \label{eq17}
\eeq
This $m_{a}$ acts as a potential energy function that keeps the satellites
from escaping.

To obtain an orthonormal basis in the space of $\psi$ one must define the
following {\em universal current} as a generalization of conserved currents
in field theories of structureless particles.
\beq
j^{\mu\alpha\beta}\equiv (i/2)\bar{\psi}\dellr{\mu}\gamma_{1}^{\alpha}
\gamma_{2}^{\beta}\psi, \label{eq18}
\eeq
where $\bar{\psi}$ is the adjoint of $\psi$ and
\beq
\bar{\psi}\dellr{\mu}\psi\equiv \bar{\psi}(\delr{\mu}-\dell{\mu})\psi \equiv
\bar{\psi}(\del^{\mu}\psi)-(\del^{\mu}\bar{\psi})\psi. \label{eq19}
\eeq
This yields the following conserved currents in each of the satellite sectors
\beq
j_{1}^{\alpha}\equiv \int j^{\mu\alpha\beta}d^{3}x_{\mu}d^{3}\xi_{2\beta},\;\;
j_{2}^{\beta}\equiv \int j^{\mu\alpha\beta}d^{3}x_{\mu}d^{3}\xi_{1\alpha},
\label{eq20}
\eeq
where terms like $d^{3}x_{\mu}$ represent four-vector hypersurface elements in
$x_{\mu}$ space. The integrations are done over arbitrary infinite spacelike 
hypersurfaces. A similar conserved current is obtained for the whole particle.
\beq
j^{\mu}\equiv \int j^{\mu\alpha\beta}d^{3}\xi_{1\alpha}d^{3}\xi_{2\beta}.
\label{eq21}
\eeq
It is straightforward to prove the following conservation equations using the
equations of motion\cite{biswas2}.
\beq
\del_{a\alpha}j_{a}^{\alpha}=0,\;\; a=1,2, \label{eq22}
\eeq
and
\beq
\del_{\mu}j^{\mu}=0. \label{eq23}
\eeq
All three conserved currents lead to the same conserved charge. It is given by
\beq
{\cal Q}\equiv \int j^{\mu\alpha\beta}d^{3}\xi_{1\alpha}d^{3}\xi_{2\beta}d^{3}x_{\mu}.
\label{eq24}
\eeq
Due to the conservation equations~\ref{eq22} and~\ref{eq23} it can be seen that the
three integrations over space-like hypersurfaces are independent of the choice of
any specific hypersurface. Hence, for convenience, we choose the $\xi_{a\alpha}$
hypersurfaces to be orthogonal to the total momentum $P$. This makes sure it is
purely spatial in the CM frame. So, we replace $d^{3}\xi_{a\alpha}$ by
$d^{3}\xi_{a}^{\bot}$ and $\gamma_{a}^{\alpha}$ by $\gamma_{a}^{\|}$ where
\beq
\gamma_{a}^{\|}\equiv -\gamma_{a\alpha}\hat{P}^{\alpha}. \label{eq25}
\eeq
For the $d^{3}x_{\mu}$ integration we choose the purely spatial components ${\bf x}$
in the laboratory frame. Hence, $\dellr{\mu}$ can be replaced by $\dellr{0}$. This
gives the conserved charge to be
\beq
{\cal Q}\equiv (-i/2)\int\bar{\psi}\dellrs{0}\gamma_{1}^{\|}\gamma_{2}^{\|}\psi
d^{3}\xi_{1}^{\bot}d^{3}\xi_{2}^{\bot}d^{3}{\bf x}. \label{eq26}
\eeq
This conserved charge suggests the following natural norm for the Hilbert space of
$\psi$.
\beq
(\psi,\psi)\equiv -i/2\int\bar{\psi}\dellrs{0}\gamma_{1}^{\|}\gamma_{2}^{\|}\psi
d^{3}\xi_{1}^{\bot}d^{3}\xi_{2}^{\bot}d^{3}{\bf x}. \label{eq27}
\eeq
This leads to the following definition of the inner product.
\beq
(\phi,\psi)\equiv -i/2\int\bar{\phi}\dellrs{0}\gamma_{1}^{\|}\gamma_{2}^{\|}\psi
d^{3}\xi_{1}^{\bot}d^{3}\xi_{2}^{\bot}d^{3}{\bf x}. \label{eq28}
\eeq
The above inner product definition allows us to identify the following
orthonormal basis for the set of solutions of the equations of motion.
\beqar
\psi_{{\bf k}E_{1}E_{2}} & \equiv & [k^{0}(2\pi)^{3}]^{-1/2}
\Psi(E_{1},E_{2},\xi_{1}^{\bot},\xi_{2}^{\bot})\cdot \nonumber \\
& & \cdot\exp[-iE_{1}\xi_{1}^{\|}-iE_{2}\xi_{2}^{\|}]\exp(ik\cdot x), \label{eq29}
\eeqar
where $k$ is the four-vector eigenvalue of the total momentum $P$, ${\bf k}$
is its three-vector part and $k^{0}$ is its zeroth component. $E_{a}$ is the
total energy of the $a$-th satellite. So it is the eigenvalue of $\pi_{a}^{\|}$.
Note that $E_{1}+E_{2}=m$, but it can be negative. This requires the usual
explanation of an antiparticle being a particle going backward in time. So the
physically measurable mass is still positive.
For $\psi_{{\bf k}E_{1}E_{2}}$ to be a solution of the equations of motion
in the satellite sectors, $\Psi(E_{1},E_{2},\xi_{1}^{\bot},\xi_{2}^{\bot})$
must satisfy the following eigenvalue equations for $a=1,2$.
\beq
H_{a}\Psi(E_{1},E_{2},\xi_{1}^{\bot},\xi_{2}^{\bot})=
E_{a}\Psi(E_{1},E_{2},\xi_{1}^{\bot},\xi_{2}^{\bot}), \label{eq30}
\eeq
where
\beq
H_{a}\equiv \alpha_{a}^{\bot}\cdot\pi_{a}^{\bot}+i\gamma_{a}^{\|}m_{a},\;\;\;
\alpha_{a}^{\bot}\equiv -\gamma_{a}^{\|}\gamma_{a}^{\bot}, \label{eq31}
\eeq
with ``$\|$'' and ``$\bot$'' superscripts given the meaning of equation~\ref{eq6}.
It is to be noted that $\Psi(E_{1},E_{2},\xi_{1}^{\bot},\xi_{2}^{\bot})$ also
depends on spin and orbital angular momentum quantum numbers due rotational symmetry.
The labels for these quantum numbers are suppressed for brevity of notation. Also,
the spectrum for $E_{a}$ is expected to be discrete as $m_{a}$ produces a confining effect.
For $\psi_{{\bf k}E_{1}E_{2}}$ to be a solution of the whole particle equation
of motion, the following must be satisfied.
\beq
k^{0}=\sqrt{{\bf k}^{2}+(E_{1}+E_{2})^{2}}. \label{eq32}
\eeq
$\Psi$ may be normalized in the usual fashion.
\beqar
\int\bar{\Psi}(E'_{1},E'_{2},\xi_{1}^{\bot},\xi_{2}^{\bot})\gamma_{1}^{\|}
\gamma_{2}^{\|}\Psi(E_{1},E_{2},\xi_{1}^{\bot},\xi_{2}^{\bot})
d^{3}\xi_{1}^{\bot}d^{3}\xi_{2}^{\bot} = && \nonumber \\
= \delta_{E'_{1}E_{1}}\delta_{E'_{2}E_{2}},\;\;\;\;\;\; &&
\label{eq33}
\eeqar
where $\delta_{E'E}$ is the Kronecker delta and, once again, the angular momentum 
labels are suppressed and understood to be included in the corresponding energy
label. Using these
conditions, it can be verified that the $\psi_{{\bf k}E_{1}E_{2}}$ are truly
orthonormal.
\beq
(\psi_{{\bf k}'E'_{1}E'_{2}},\psi_{{\bf k}E_{1}E_{2}})=
\delta_{E'_{1}E_{1}}\delta_{E'_{2}E_{2}}\delta({\bf k}'-{\bf k}), \label{eq34}
\eeq
where $\delta({\bf k}'-{\bf k})$ is the Dirac delta.

Now we are ready for second quantization. The standard prescription for
canonical quantization will be used. However, it is critical to note that
individual satellites and the vertex are not second quantized. It is the whole
particle wavefunction $\psi$ that is second quantized. The energies and
angular momenta of the satellites are treated as extra degrees of freedom
(quantum numbers) of the whole particle wavefunction.

First, a Lagrangian for the particle field is defined as follows.
\beq
{\cal L}=-\int\bar{\psi}\gamma_{1}^{\|}\gamma_{2}^{\|}[\dells{\mu}\delr{\mu}+m^{2}]\psi
d^{3}\xi_{1}^{\bot}d^{3}\xi_{2}^{\bot}, \label{eq35}
\eeq
where $m$ is given by equation~\ref{eq14}. The momentum conjugate to $\psi$ would
then be
\beq
\phi\equiv\frac{\del{\cal L}}{\del(\del_{0}\psi)}=-\bar{\psi}\gamma_{1}^{\|}\gamma_{2}^{\|}
\dell{0}=\del^{0}\psi^{\dagger}, \label{eq36}
\eeq
where the Hermitian adjoint $\psi^{\dagger}$ is related to the adjoint $\bar{\psi}$
as follows.
\beq
\psi^{\dagger}=\bar{\psi}(i\gamma_{1}^{\|})(i\gamma_{2}^{\|}). \label{eq37}
\eeq
Then, from the spin-statistics theorem, one concludes that the second quantization condition
can be written symbolically as the following equal-time commutator.
\beq
[\psi,\phi]=i\delta, \label{eq38}
\eeq
where the $\delta$ is a delta function over all degrees of freedom.

Now, $\psi$ can be expanded in terms of the basis set of equation~\ref{eq29} as follows.
\beqar
\psi & = & \int d^{3}{\bf k}\sum_{E_{1}E_{2}}[2k^{0}(2\pi)^{3}]^{-1/2}
\Psi(E_{1},E_{2},\xi_{1}^{\bot},\xi_{2}^{\bot})\cdot \nonumber \\
&&\cdot\exp[-iE_{1}\xi_{1}^{\|}-iE_{2}\xi_{2}^{\|}]\cdot \nonumber \\
& & \cdot[b({\bf k},E_{1},E_{2})\exp(ik\cdot x)+\nonumber \\
&& +d^{*}({\bf k},E_{1},E_{2})\exp(-ik\cdot x)].
\label{eq39}
\eeqar
As in usual field theories, the $b$ and $d^{*}$ coefficients are used to separate particle
and antiparticle states. $d^{*}$ represents the Hermitian adjoint of $d$ in a field operator
sense. Then, the quantization condition of equation~\ref{eq38}
reduces to the following (as before, the energy labels are understood to include
angular momentum labels).
\beqar
&& [b({\bf k},E_{1},E_{2}),b^{*}({\bf k}',E'_{1},E'_{2})] \nonumber \\
&&\;\;\;\; = -[d^{*}({\bf k},E_{1},E_{2}),d({\bf k}',E'_{1},E'_{2})] \nonumber \\
&&\;\;\;\; = \delta^{3}({\bf k}-{\bf k}')\delta_{E'_{1}E_{1}}\delta_{E'_{2}E_{2}},
\label{eq40}
\eeqar
and all other commutators of $b$, $b^{*}$, $d$ and $d^{*}$ vanish. This allows the
building of the usual Fock space with $b^{*}$ being the particle creation operator,
$b$ the particle annihilation operator, $d^{*}$ the antiparticle creation operator
and $d$ the antiparticle annihilation operator. The necessary
vacuum state can be shown to be stable\cite{biswas2}.

\section{The structure of fermions}
\begin{figure}
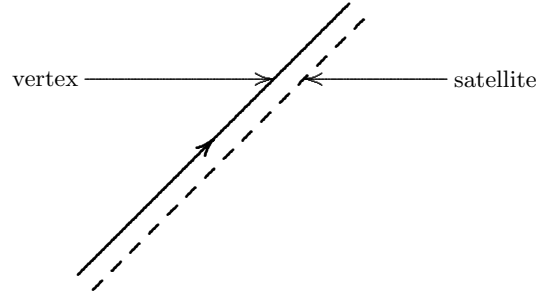

\beginpicture
\setcoordinatesystem units <1mm,1mm>
\setplotarea x from -25 to 25, y from -20 to 20
\setplotsymbol ({.})
\setdashes
\plot -18 -18  18 18 /
\setsolid
\arrow <7pt> [.2,.67] from -20 -16 to -2 2
\plot -2 2  16 20 /
\setplotsymbol ({\fiverm .})
\put {vertex} [r] at -20 10
\arrow <7pt> [.2,.67] from -19 10 to 6 10
\put {satellite} [l] at 30 10
\arrow <7pt> [.2,.67] from 29 10 to 10 10
\endpicture
\caption{Model of a fermion \label{fig2}}
\end{figure}

A fermionic particle is also modeled as a composite. However, for this composite, 
the vertex is spin-half instead of spin-zero and there is only one satellite which is
spin-zero instead of spin-half. In Feynman diagrams it will be represented as shown 
in figure~\ref{fig2}.
The coordinates for the vertex and the satellite can be chosen in a fashion similar
to bosons. The phase space $S_{p}$ is as follows.
\beq
S_{p}= \{P,\pi^{\|},\pi^{\bot},x,\xi^{\|},\xi^{\bot}\}, \label{eq41}
\eeq
where $P$ and $x$ are the momentum and position of the whole particle
and $\pi$ and $\xi$ are the momentum and position of the spin-zero satellite. The
non-zero commutators are as follows.
\beq
[x,P]=i\eta,\;\;[\xi^{\|},\pi^{\|}]=-i,\;\;
[\xi^{\bot},\pi^{\bot}]=iP^{\bot}. \label{eq42}
\eeq
The wavefunction $\psi$ is a function of a maximal mutually commuting subset of
$S_{p}$ as follows.
\beq
\psi = \psi(x,\xi^{\|},\xi^{\bot}). \label{eq43}
\eeq
The commutation conditions of equation~\ref{eq42} lead to the following
differential operator representation of the momenta.
\beq
i(-\pi^{\|},\pi^{\bot})=\del_{1\alpha}\equiv
\left(\frac{\del}{\del\xi^{\|}},\nabla\right), \label{eq44}
\eeq
and
\beq
iP_{\alpha}=\del_{\alpha}\equiv\frac{\del}{\del x^{\alpha}}. \label{eq45}
\eeq
$\nabla$ is the gradient operator in the three-vector space of $\xi^{\bot}$.
The spin-half vertex equation of motion written as the whole particle equation
should be akin to the Dirac equation. So, it is chosen to be as follows.
\beq
D_{0}\psi\equiv(\gamma^{\alpha}\del_{\alpha}+m)\psi=0, \label{eq46}
\eeq
where, as before, the CM energy of the vertex is chosen to vanish and hence
the rest mass of the whole particle becomes the CM energy of the satellite.
\beq
m=\pi^{\|}. \label{eq47}
\eeq
The spin-zero satellite has an equation of motion as follows.
\beq
D_{1}\psi\equiv(\del_{1\mu}\del_{1}^{\mu}-m_{1}^{2})\psi=0, \label{eq48}
\eeq
where $m_{1}$ includes a confining potential for the satellite. So,
\beq
m_{1}=m_{1}(\xi^{\bot}). \label{eq49}
\eeq
This form of $m_{1}$ ascertains the consistency of the equations of
motion as seen in equation~\ref{eq16}.

The universal current for fermions would then be given by
\beq
j^{\mu\alpha}\equiv (1/2)\bar{\psi}\dellrss{1}{\mu}\gamma^{\alpha}\psi. \label{eq50}
\eeq
This helps define the conserved charge as
\beq
{\cal Q}\equiv \int j^{\mu\alpha}d^{3}\xi_{\mu}d^{3}x_{\alpha}
=(1/2)\int\bar{\psi}\dellrss{1}{\|}\gamma^{0}\psi d^{3}\xi^{\bot}d^{3}{\bf x}.
\label{eq51}
\eeq
Then the inner product is seen to be
\beq
(\phi,\psi)\equiv 1/2\int\bar{\phi}\dellrss{1}{\|}\gamma^{0}\psi
d^{3}\xi^{\bot}d^{3}{\bf x}. \label{eq52}
\eeq
This allows the following orthonormal basis for the solution set of the
equations motion.
\beq
\psi^{+}_{{\bf k}Er}\equiv\sqrt{\frac{m}{Ek^{0}(2\pi)^{3}}}
\Psi(E,\xi^{\bot})\exp[-iE\xi^{\|}]u_{r}(k)\exp(ik\cdot x), \label{eq53}
\eeq
for positive energy states and 
\beq
\psi^{-}_{{\bf k}Er}\equiv\sqrt{\frac{m}{Ek^{0}(2\pi)^{3}}}
\Psi(E,\xi^{\bot})\exp[-iE\xi^{\|}]v_{r}(k)\exp(-ik\cdot x), \label{eq54}
\eeq
for negative energy states. Here $u_{r}(k)$ and $v_{r}(k)$ are the
usual free-field Dirac equation solutions with $r=\pm$ giving the two
possible spin states. It is to be noted from equation~\ref{eq47} that
$m=E$, as $E$ is an eigenvalue of $\pi^{\|}$. Hence, the normalization
factor above simplifies to $((2\pi)^{3}k_{0})^{-1/2}$. Also, $k_{0}$ is found
to be
\beq
k^{0}=\sqrt{{\bf k}^{2}+E^{2}}, \label{eq55}
\eeq
and $\Psi(E,\xi^{\bot})$ must satisfy the following eigenvalue equation.
\beq
H\Psi(E,\xi^{\bot})=E\Psi(E,\xi^{\bot}), \label{eq56}
\eeq
where
\beq
H\equiv \sqrt{(\pi^{\bot})^{2}+m_{1}^{2}}. \label{eq57}
\eeq
$\Psi$ can be normalized as follows.
\beq
\int\Psi^{*}(E',\xi^{\bot})\Psi(E,\xi^{\bot})d^{3}\xi^{\bot}=\delta_{E'E}. \label{eq58}
\eeq

Now we are ready to second quantize these fermion fields. The Lagrangian is
\beqar
{\cal L} & = & -(1/2)\int\bar{\psi}\dellrss{1}{\|}[\gamma^{\alpha}\del_{\alpha}+m]\psi
d^{3}\xi^{\bot} \nonumber \\
& = & \int\bar{\psi}\dellss{1}{\|}[\gamma^{\alpha}\del_{\alpha}+m]\psi
d^{3}\xi^{\bot}. \label{eq59}
\eeqar
So the momentum conjugate to $\psi$ is
\beq
\phi=\frac{\del{\cal L}}{\del(\del_{0}\psi)}=
\del_{1}^{\|}\bar{\psi}\gamma^{0}=-i\del_{1}^{\|}\psi^{\dagger}. \label{eq60}
\eeq
Symbolically, the second quantization condition is given by the following
anticommutator.
\beq
\{\psi,\phi\}=i\delta. \label{eq61}
\eeq
All other anticommutators are zero.
Anticommutators are used here in keeping with the spin-statistics theorem.
In terms of creation and annihilation operators this relation can be written as
\beqar
\{b({\bf k},E,r),b^{*}({\bf k}',E',r')\} & = &
\{d({\bf k},E,r),d^{*}({\bf k}',E',r')\}= \nonumber \\
& = &\delta^{3}({\bf k}-{\bf k}')\delta_{E'E}\delta_{r'r},
\label{eq62}
\eeqar
where $b$ and $d$ are defined as the following expansion coefficients of
$\psi$ in terms of the orthonormal basis.
\beq
\psi = \int d^{3}{\bf k}\sum_{E}\sum_{r}[b({\bf k},E,r)\psi^{+}_{{\bf k}Er}+
d^{*}({\bf k},E,r)\psi^{-}_{{\bf k}Er}]. \label{eq63}
\eeq
All anticommutators of $b$, $b^{*}$, $d$ and $d^{*}$ other than the ones
in equation~\ref{eq62} vanish.

\section{Interaction centers}
The composite particle models described in the last two sections can be
generalized further. One may have a spin-zero vertex with any number of
spin-half as well as spin-zero satellites. Similarly, a spin-half vertex
may have any number of spin-zero or spin-half satellites.
As a result, the photon, the $W^{\pm}$, the $Z$ and all mesons can
be modelled as bosonic composites while the electron,
the muon, the tau, all neutrinos and all baryons can be modelled as fermionic composites.

However, at present, I shall consider only electromagnetic interactions. The
photon is a composite boson and the electron a composite fermion. The fermionic
satellites of a free photon will have energies that are equal in magnitude but
opposite in sign. This gives the photon a zero rest mass.
The following interaction Lagrangian can describe all experimentally 
observed interactions.
\beqar
{\cal L}_{I} & = & e\int \bar{\psi}_{\bar{a}}(x,\xi_{1})
[d^{3}\xi_{1}^{\bot}\dellss{1}{\|}]
 \gamma_{a\bar{b}}^{\mu}\bar{\phi}_{b\bar{c}}(x,\xi_{1},\xi_{2})\cdot \nonumber \\
&& \cdot\gamma_{c\bar{d}\mu}
[d^{3}\xi_{2}^{\bot}\delrss{2}{\|}]\psi_{d}(x,\xi_{2}), \label{eq64}
\eeqar
where $\psi$ is used to represent fermion fields and $\phi$ for boson fields.
The spinor indices are written explicitly and indices like $\bar{a}$ are used to label
adjoint spinor indices. Indices like $a$ and $\bar{a}$ used here are not to be confused with the satellite number index used in section~\ref{boson}. Contractions are over $a$ and
$\bar{a}$, $b$ and $\bar{b}$ etc.. $\bar{\phi}$ is the adjoint of $\phi$ for only 
the second satellite sector. Hence, $\bar{c}$ is used to denote the second spinor index.
$e$ is the electron charge. The integration variables $\xi_{1}^{\bot}$ and $\xi_{2}^{\bot}$
are coordinates orthogonal to the corresponding {\em electron} momenta and not the photon
momentum.

\begin{figure}
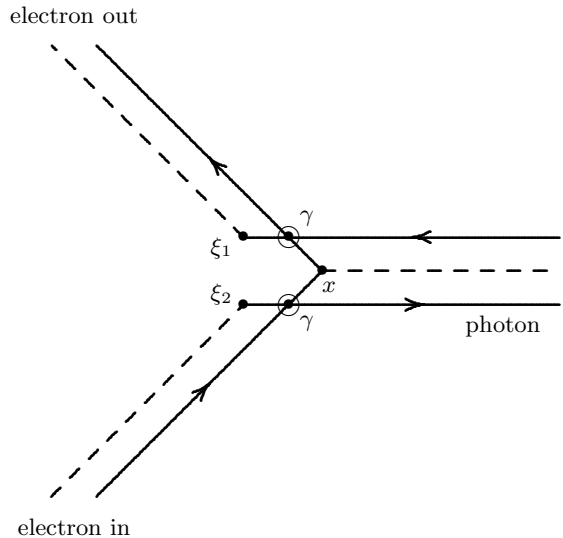

\beginpicture
\setcoordinatesystem units <1.5mm,1.5mm>
\setplotarea x from -25 to 25, y from -25 to 25
\setplotsymbol ({.})
\setdashes
\plot -20 -20  -3 -3 /
\plot -3 3  -20 20 /
\plot 4 0  25 0 /
\setsolid
\arrow <7pt> [.2,.67] from -16 -20 to -6 -10
\plot -6 -10  4 0 /
\arrow <7pt> [.2,.67] from 4 0 to -6 10
\plot -6 10  -16 20 /
\arrow <7pt> [.2,.67] from -3 -3 to 13 -3
\plot 13 -3  25 -3 /
\arrow <7pt> [.2,.67] from 25 3 to 12 3
\plot 12 3  -3 3 /
\setplotsymbol ({\fiverm .})
\put {$\xi_{1}$} [tr] at -4 3
\put {$\xi_{2}$} [br] at -4 -3
\put {$x$} [lt] at 4 -1
\put {$\gamma$} [bl] at 2 4
\put {$\gamma$} [tl] at 2 -4
\put {$\bullet$} at -3 -3
\put {$\bullet$} at -3 3
\put {$\bullet$} at 4 0
\put {$\bigcirc$} at 1 -3
\put {$\bigcirc$} at 1 3
\put {$\bullet$} at 1 -3
\put {$\bullet$} at 1 3
\put {electron in} [t] at -18 -22
\put {electron out} [b] at -18 22
\put {photon} [t] at 20 -4
\endpicture
\caption{Interaction center for QED. \label{fig3}}
\end{figure}

In Feynman diagrams the effects of ${\cal L}_{I}$ will be called the interaction {\em centers}
instead of {\em vertices} as they are usually called. This is because
the name {\em vertex} has been used for a component of a composite particle.
Figure~\ref{fig3} shows the diagram for such an interaction center. Note that the
diagram shows a coupling of the coordinates of the two satellites of the photon to the
satellites of the two electrons. This accurately describes ${\cal L}_{I}$ as shown
in equation~\ref{eq64}. Equation~\ref{eq64} also shows a coupling of spins
of each photon satellite to each of the two electron vertices. They are mediated
by the two $\gamma_{\mu}$ matrices. This is represented in the diagram by the intersections 
of the fermion lines. In usual field theories, it is known that fermion lines do not begin
or end at an interaction center. Here, that is still true. However, little pieces of the
fermion lines extend out from the intersection points (marked by circles) to represent
their couplings to bosonic components. One way of visualizing this is
to consider the photon to be forming a fermionic current due to its fermionic satellites.
It is given by
\beq
A_{b\bar{d}\mu}(x,\xi_{1},\xi_{2})=\bar{\phi}_{b\bar{c}}(x,\xi_{1},\xi_{2})\gamma_{c\bar{d}\mu}.
\label{eq65}
\eeq
The current of the electron fields would be
\beq
J_{d\bar{b}}^{\mu}(x,\xi_{1},\xi_{2}) = 
\bar{\psi}_{\bar{a}}(x,\xi_{1})\dellss{1}{\|}\gamma_{a\bar{b}}^{\mu}
\delrss{2}{\|}\psi_{d}(x,\xi_{2}). \label{eq66}
\eeq
Then the interaction can be seen as an interaction of two currents:
\beq
{\cal L}_{I}=e\int J_{d\bar{b}}^{\mu}(x,\xi_{1},\xi_{2})A_{b\bar{d}\mu}(x,\xi_{1},\xi_{2})
d^{3}\xi_{1}^{\bot}d^{3}\xi_{2}^{\bot}. \label{eq67}
\eeq
At the same time, it is also noticed that $A_{b\bar{d}\mu}(x,\xi_{1},\xi_{2})$ is the composite
particle analog of the usual vector field $A_{\mu}(x)$ representing photons.

At first sight, the above choice of an interaction center may seem arbitrary. However,
closer inspection shows that there aren't too many other choices. Some of the natural
restrictions on ${\cal L}_{I}$ are as follows.
\begin{itemize}
\item It must be Lorentz invariant.
\item It must be a natural extension of the free field Lagrangians of both photons and electrons.
\item It must be symmetric in the two electron lines.
\item It must agree with all experimentally verified results of usual QED.
\item No fermionic line (satellites or vertex) can end at the interaction center.
\item The photon field should not appear as a pure gauge.
\end{itemize}
The following is a slight variation of ${\cal L}_{I}$ that satisfies all but one of the above
restrictions.
\beqar
{\cal L}_{I0} & = & e\int \bar{\psi}_{\bar{a}}(x,\xi_{1})
[d^{3}\xi_{1}^{\bot}\dellss{1}{\|}]
 \gamma_{a\bar{b}}^{\mu}\bar{\phi}_{c\bar{d}}(x,\xi_{1},\xi_{2})\cdot \nonumber \\
&& \cdot\gamma_{d\bar{c}\mu}
[d^{3}\xi_{2}^{\bot}\delrss{2}{\|}]\psi_{b}(x,\xi_{2}). \label{eq68}
\eeqar
Here the spinor indices of the photon satellites couple with each other and the two electron
spins couple with each other. This form of the interaction will not be discussed any further here
as it does not produce the experimentally observed anomalous electron magnetic moment. It only
produces a {\em finite} charge renormalization. Hence, it can be ignored.

\section{Perturbation theory and Haag's theorem}
Using the interaction described by equation~\ref{eq64} (or~\ref{eq68}) one can find scattering
matrix elements for QED interactions. However, only tree diagrams need be considered. The
effects of closed loops in diagrams are completely included in the interaction center. So,
instead of infinite loop integrals we will have integrals over internal coordinates like
$\xi_{1}^{\bot}$ and $\xi_{2}^{\bot}$. If the field operators are expanded in terms of the
basis sets (equations~\ref{eq29}, \ref{eq53} and~\ref{eq54}), these integrals will produce
summations over the bound state energies (and angular momenta) of the satellites. 
As these energies are discrete,
the summations are likely to produce finite results. This is qualitatively similar to Planck's
derivation of the blackbody radiation spectrum -- a summation over discrete energies
produced the correct finite result instead of the infinite result of Raleigh-Jeans obtained
by integration over continuous energies.

Physically, this makes sense as the virtual particles of loop diagrams are like internal 
bound particles. So, their states should be expected to display a discrete spectrum. But
in traditional QED there is no way of recognizing this feature and hence, it produces
infinite integrals for closed loops in diagrams. In the composite particle
QED formulated here, this discreteness appears naturally and hence, it is likely to produce
finite results.

However, strict adherence to perturbation theory requires the use of the interaction picture.
This will still produce closed loop diagrams of composite particles. So, here we need to
deviate from the interaction picture as follows.

\begin{quotation}
{\em Of all possible Feynman diagrams produced by the interaction picture, only
the tree diagrams will be considered physical.}
\end{quotation}

This assumption does not weaken the theory in any way as the effects of the closed loop
diagrams are included in the internal structure. At the same time, it avoids infinite
renormalizations. It also has an unintended, but highly desirable, side effect -- Haag's
theorem\cite{wightman} is no longer an impediment. Let $U_{I}$ be the usual unitary
operator that transforms free fields to interacting fields. In the present formulation,
we are keeping only tree diagrams. This means we are keeping only a part of $U_{I}$ --
say $V_{I}$. This $V_{I}$ will not be unitary and hence, Haag's theorem is no longer
valid in this formulation.

\section{QED -- agreement with experiment}
To find experimental consequences of the present formulation, we shall compute
the magnetic moment of the electron. For lowest order effects we compute the interaction
energy of an electron and a photon field while all satellites are in their ground
states. Let the incoming electron have a momentum $p$, the outgoing electron a momentum
$p'$ and the incoming photon a momentum $q$. Then, integrating ${\cal L}_{I}$ of 
equation~\ref{eq64} over $d^{4}x$ produces the Dirac delta $\delta(p'-p-q)$. Hence,
\beq
q=p'-p. \label{eq69}
\eeq
The remaining part of the integral is the interaction energy of interest. It is
\beq
H_{I}=\bar{u}_{\bar{a}}(p')\gamma^{\mu}_{a\bar{b}}
W_{b}\bar{W}_{\bar{c}}\gamma_{c\bar{d}\mu}u_{d}(p). \label{eq70}
\eeq
where $\bar{u}_{\bar{a}}(p')$, and $u_{d}(p)$ are Dirac spinor components for 
the outgoing and incoming free electrons. Also,
\beqar
W_{b} & = & \int \Psi^{*}(E_{0},\xi_{1}^{\bot})
\Psi_{b}(e_{0},\xi_{1}^{\bot^{\prime}})d^{3}\xi_{1}^{\bot},
\label{eq71} \\
\bar{W}_{\bar{c}} & = & \int \Psi(E_{0},\xi_{2}^{\bot})
\bar{\Psi}_{\bar{c}}(e_{0},\xi_{2}^{\bot^{\prime}})d^{3}\xi_{2}^{\bot}.
\label{eq72}
\eeqar
Here $E_{0}$ is the ground state energy of the electron satellite and $e_{0}$ 
that of each of the two
photon satellites. $\Psi$ is the electron satellite ground state wavefunction and
$\Psi_{b}$ the ground state wavefunction of one of the two photon satellites. $\Psi_{b}$
is obtained by separating the photon $\Psi$ of equation~\ref{eq30} into the two
separate satellite wavefunctions. The $\xi_{1}^{\bot}$ are components
of $\xi_{1}$ in a hypersurface orthogonal to the corresponding electron momentum. The
$\xi_{1}^{\bot^{\prime}}$ are components of $\xi_{1}$ in a hypersurface orthogonal to the corresponding photon momentum. To compute the integrals of equations~\ref{eq71} and~\ref{eq72},
one needs the mass functions of equations~\ref{eq17} and~\ref{eq49}. Explicit forms
for these functions will be introduced in a later article. 
For now, it is to be noticed that one may write
\beq
W_{b}\bar{W}_{\bar{c}}\gamma_{c\bar{d}\mu}=(I_{b\bar{d}}+\Delta_{b\bar{d}})a_{\mu}(q),
\label{eq73}
\eeq
where $I_{b\bar{d}}$ is the identity for Dirac spinors and $a_{\mu}(q)$ can be interpretted
as the momentum representation of the usual electromagnetic vector potential for a momentum
$q$. The $\Delta_{b\bar{d}}$ provides the anomalous magnetic moment. For this to agree with
experiment in lowest order one must have
\beq
\Delta_{b\bar{d}} = \frac{i\alpha}{4\pi E_{0}}\gamma_{b\bar{d}\nu}q^{\nu}, \label{eq74}
\eeq
where $\alpha$ is the fine structure constant. A complete computation of
$\Delta_{b\bar{d}}$ will be presented in a later article.

\section{Conclusion}
A quantum field theory of composite electrons and photons is suggested. It is distinctly
different from string theories in many ways - in particular, it does not require space-time
to have dimensions greater than four. It also does not require infinite renormalizations.
As a side-effect, the no-interaction result of Haag's theorem is avoided as well.
Experimentally testable aspects of usual QED (anomalous electron magnetic moment, Lamb shift etc.)
are expected to be reproduced in this theory.


\begin{thebibliography}{99}
\bibitem{sudarshan}D.~G.~Currie, T.~F.~Jordan and E.~C.~G.~Sudarshan, 
\textit{Rev.~Mod.~Phys.,} {\bf 38}, 350 (1963).
\bibitem{arens}R.~Arens, \textit{Nuovo Cimento B,} {\bf 21}, 395 (1975).
\bibitem{droz}Ph.~Droz-Vincent, \textit{Rep.~Math.~Phys.,} {\bf 8}, 79 (1975).
\bibitem{todorov}V.~V.~Molotov and I.~T.~Todorov, JINR Report EZ-12270, Dubna (1979).
\bibitem{komar}A.~Komar, \textit{Phys.~Rev.~D,} {\bf 18}, 1881, 1887 (1978).
\bibitem{biswas1}T.~Biswas, \textit{Nuovo Cimento A,} {\bf 88}, 154 (1985).
\bibitem{biswas2}T.~Biswas, \textit{Nuovo Cimento A,} {\bf 107}, 863 (1994).
\bibitem{wightman}R.~F.~Streater and A.~S.~Wightman, \textit{PCT, Spin and Statistics, and All That,}
pp.~165-166 (Benjamin/Cummings, 1980).
\end{thebibliography}
\end{document}